\documentclass[12pt]{iopart}
\expandafter\let\csname equation*\endcsname\relax
\expandafter\let\csname endequation*\endcsname\relax
\usepackage{amsfonts}
\usepackage{amsmath}
\usepackage{mathtools}
\usepackage{xcolor}
\usepackage{graphicx}
\usepackage{placeins}
\usepackage[backend=biber, sorting=none]{biblatex}
\usepackage{microtype}
\usepackage{booktabs}       % professional-quality tables
\usepackage{tabularx}
\usepackage{multirow}
\usepackage{caption}
\usepackage{subcaption}
\renewcommand{\vec}[1]{{\boldsymbol{#1}}}
\newcommand{\hf}{\text{HF}}
\newcommand{\nel}{{n_\text{el}}}
\newcommand{\nup}{{n_\uparrow}}

\newcommand{\ndet}{{N_\text{det}}}
\newcommand{\nnuc}{{N_\text{nuc}}}
\newcommand{\norb}{{N_\text{orb}}}
\newcommand{\nbasis}{{N_\text{basis}}}
\newcommand{\embdim}{{D_\text{emb}}}
\newcommand{\sumions}{\sum_{I=1}^{\nnuc}}

\begin{document}
\title{Towards a Foundation Model for Neural Network Wavefunctions}
\author{Michael Scherbela$^{1, 4}$, Leon Gerard$^{2, 4}$, Philipp Grohs$^{1, 2, 3}$}
\address{$^1$ Faculty of Mathematics, University of Vienna, Austria}
\address{$^2$ Research Network Data Science, University of Vienna, Austria}
\address{$^3$ Johann Radon Institute for Computational and Applied Mathematics,  Austrian Academy of Sciences, Austria}
\address{$^4$ Equal contribution}
\ead{philipp.grohs@univie.ac.at}

\begin{abstract}
Deep neural networks have become a highly accurate and powerful wavefunction ansatz in combination with variational Monte Carlo methods for solving the electronic Schrödinger equation. 
However, despite their success and favorable scaling, these methods are still computationally too costly for wide adoption.
A significant obstacle is the requirement to optimize the wavefunction from scratch for each new system, thus requiring long optimization.
In this work, we propose a novel neural network ansatz, which effectively maps uncorrelated, computationally cheap Hartree-Fock orbitals, to correlated, high-accuracy neural network orbitals.
This ansatz is inherently capable of learning a single wavefunction across multiple compounds and geometries, as we demonstrate by successfully transferring a  wavefunction model pre-trained on smaller fragments to larger compounds. 
Furthermore, we provide ample experimental evidence to support the idea that extensive pre-training of a such a generalized wavefunction model across different compounds and geometries could lead to a foundation wavefunction model. 
Such a model could yield high-accuracy ab-initio energies using only minimal computational effort for fine-tuning and evaluation of observables. 
\end{abstract}
\maketitle
\ioptwocol

\section{Introduction}
Accurate predictions of quantum mechanical properties for molecules is of utmost importance for the development of new compounds, such as catalysts, or pharmaceuticals. 
For each molecule the solution to the Schrödinger equation yields the wavefunction and electron density, and thus in principle gives complete access to its chemical properties. 
However, due to the curse of dimensionality, computing accurate approximations to the Schrödinger equation quickly becomes computationally intractable with increasing number of particles.
Recently, deep-learning-based Variational Monte Carlo (DL-VMC) methods have emerged as a high-accuracy approach with favorable scaling $\mathcal{O}(N^4)$ in the number of particles $N$ \cite{hermannDeepneuralnetworkSolutionElectronic2020}. 
These methods use a deep neural network as ansatz for the high-dimensional wavefunction, and minimize the energy of this ansatz to obtain the ground-state wavefunction.
Based on two major architectures for the treatment of molecules in first quantization, PauliNet \cite{hermannDeepneuralnetworkSolutionElectronic2020} and FermiNet \cite{pfauInitioSolutionManyelectron2020}, several improvements and applications have emerged.
On the one hand, enhancements of architecture, optimization and overall approach have led to substantial improvements in accuracy or computational cost \cite{spencerBetterFasterFermionic2020, vonglehnSelfAttentionAnsatzAbinitio2022, gerardGoldstandardSolutionsSchrodinger2022, ferminet_dmc_bytedance, FermiNet_DMC}.
On the other hand, these methods have been adapted to many different systems and observables: model systems of solids \cite{cassella_model_solids_physrevlett_2023, Wilson_HEG2022}, real solids \cite{li_ab_solids_2022}, energies and properties of individual molecules \cite{ hermannDeepneuralnetworkSolutionElectronic2020, pfauInitioSolutionManyelectron2020, Han_2019, gerardGoldstandardSolutionsSchrodinger2022}, forces \cite{qianYubing_dl_forces_2022, scherbelaSolvingElectronicSchrodinger2022}, excited states \cite{entwistle_electronic_2023} and potential energy surfaces \cite{scherbelaSolvingElectronicSchrodinger2022, gaoAbInitioPotentialEnergy2021, gao-pesnet++_2022}.
Furthermore, similar methods have been developed and successfully applied to Hamiltonians in second quantization \cite{carleoSolvingQuantumManybody2017, kochkovVariationalOptimizationAI2018}.

We want to emphasize that DL-VMC is an ab-initio method, that does not require any input beyond the Hamiltonian, which is defined by the molecular geometry.
This differentiates it from surrogate models, which are trained on results from ab-initio methods to either predict wavefunctions \cite{schuttUnifyingMachineLearning2019, unkeSEEquivariantPrediction2021} or observables \cite{batatiaDesignSpaceEquivariant2022}.

Despite the improvements in DL-VMC, it has not yet been widely adopted, in part due to the high computational cost.
While DL-VMC offers favorable scaling, the method suffers from a 
large prefactor, caused by an expensive optimization with potentially slow convergence towards accurate approximations. 
Furthermore this optimization needs to be repeated for every new system, leading to prohibitively high computational cost for large-scale use.
This can be partially overcome by sharing a single ansatz with identical parameters across different geometries of a compound, allowing more efficient computation of Potential Energy Surfaces (PES) \cite{scherbelaSolvingElectronicSchrodinger2022, gaoAbInitioPotentialEnergy2021, gao-pesnet++_2022}. 
However, these approaches have been limited to different geometries of a single compound and do not allow successful transfer to new compounds.
A key reason for this limitation is that current architectures explicitly depend on the number of orbitals (and thus electrons) in a molecule. 
Besides potential generalization issues, this prevents a transfer of weights between different compounds already by the fact that the shapes of weight matrices are different for compounds of different size.

In this work we propose a novel neural network ansatz, which does not depend explicitly on the number of particles, allowing to optimize wavefunctions across multiple compounds with multiple different geometric conformations. 
We find, that our model exhibits strong generalization when transferring weights from small molecules to larger, similar molecules. In particular we find that our method achieves high accuracy for the important task of relative energies.
Inspired by the success of foundation models in language \cite{gpt3_2020} or vision \cite{radfordLearningTransferableVisual2021, yuanFlorenceNewFoundation2021} -- which achieve high accuracy with minimal fine-tuning of an extensively pre-trained base-model -- we train a first base-model for neural network wavefunctions.

We evaluate our pre-trained wavefunction model by performing few-shot predictions on chemically similar molecules (in-distribution) and disparate molecules (out-of-distribution).
We find that our ansatz outperforms conventional high-accuracy methods such as CCSD(T)-ccpVTZ and that fine-tuning our pre-trained model reaches this accuracy $\approx$20x faster, than optimizing a new model.
When analyzing the accuracy as a function of pre-training resources, we find that results systematically and substantially improve by scaling up either the model size, data size or number of pre-training steps.
These results could pave the way towards a foundation wavefunction model, to obtain high-accuracy ab-initio results of quantum mechanical properties using only minimal computational effort for fine-tuning and evaluation of observables.

Additionally we compare our results to GLOBE, a concurrent work \cite{gaoGeneralizingNeuralWave2023}, which proposes reparameterization of the wavefunction based on machine-learned, localized molecular orbitals.
We find that our method in comparison achieves lower absolute energies, higher accuracy of relative energies and is better able to generalize across chemically different compounds.

\section{Results}
In the following, we briefly outline our approach and how it extends existing work in Sec. \ref{sec:extending_ansatz_to_multi_compound}.
We show the fundamental properties of our ansatz such as extensivity (Sec. \ref{sec:size_consistency}) and equivariance with respect to the sign of reference orbitals (Sec. \ref{sec:results_hf_phase}). 
We demonstrate the transferability of the ansatz when pre-training on small molecules and re-using it on larger, chemically similar molecules. 
We also compare its performance against GLOBE, a concurrent pre-print \cite{gaoGeneralizingNeuralWave2023} in Sec. \ref{sec:reuse_from_smaller}. 
Lastly, we present a first wavefunction base-model pre-trained on a large diverse dataset of 360 geometries and evaluate its downstream performance in Sec. \ref{sec:foundation_model}.

\subsection{A multi-compound wavefunction ansatz}
\label{sec:extending_ansatz_to_multi_compound}
Existing high-accuracy ansätze for neural network wavefunctions all exhibit the following structure:
\begin{align}
    &\vec{h}_i = h_\theta(\vec{r}_i, \lbrace\vec{r}\rbrace, \vec{R}, \vec{Z})  \label{eq:ferminet_embedding}\\
    &\Phi^d_{ik} =  \varphi_{dk}(\vec{r}_i) \sum_{\nu=1}^\embdim F^d_{k\nu} h_{i\nu}\label{eq:ferminet_orbitals}\\
    &\psi = \sum_{d=1}^\ndet \det\left[\Phi^d_{ik}\right]_{i,k=1\dots \nel}\label{eq:fermint_wf}
\end{align}
Eq. \ref{eq:ferminet_embedding} computes a $\embdim$-dimensional embedding of electron $i$, by taking in information of all other particles, e.g. by using attention or message passing.
Eq. \ref{eq:ferminet_orbitals} maps these high-dimensional embeddings onto $\nel \times \ndet$ orbitals (indexed by $k$), using trainable backflow matrices $\vec{F}^d$ and typically trainable envelope functions $\varphi_{dk}$.
Eq. \ref{eq:fermint_wf} evaluates the final wavefunction $\psi$ as a sum of (Slater-)determinants of these orbitals, to ensure anti-symmetry with respect to permutation of electrons.

\begin{figure}
    \centering
    \includegraphics[width=\columnwidth]{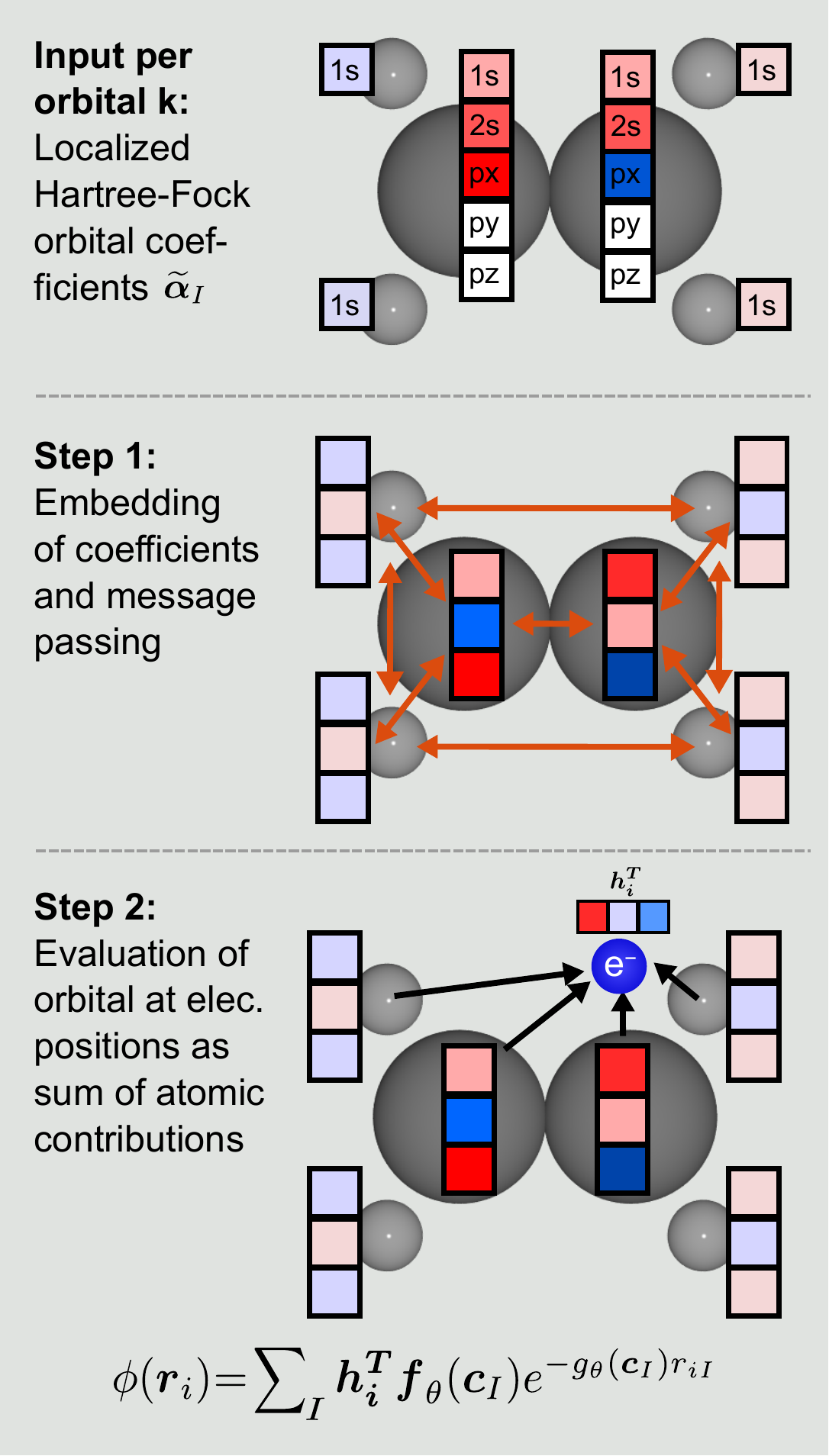}
    \caption{Illustration of the Transferrable Atomic Orbitals, demonstrated on the C=C-bond of Ethene.}
    \label{fig:architecture}
\end{figure}

While this approach works well for the wavefunctions of a single compound, it is fundamentally unsuited to represent wavefunctions for multiple different compounds at once.
The main problem lies in the matrices $\vec{F}^d$, which are of shape $[\nel \times \embdim]$, and thus explicitly depend on the number of electrons.
There are several potential options, how this challenge could be overcome. 
A na\"ive approach would be to generate a fixed number of $N_\text{orb} > \nel$ orbitals and truncate the output to the required number of orbitals $\nel$, which may differ across molecules. 
While simple to implement, this approach is however fundamentally limited to molecules with $\nel \leq N_\text{orb}$.
Another approach is to use separate matrices $\vec{F}^d_\mathcal{G}$ for each molecule or geometry $\mathcal{G}$, as was done in \cite{scherbelaSolvingElectronicSchrodinger2022}, but also this approach can fundamentally not represent wavefunctions for molecules that are larger than the ones found in the training set.
A third approach would be to not generate all orbitals in a single pass, but generate the orbitals sequentially in an auto-regressive manner, by conditioning each orbital on the previously generated orbitals. 
While this approach has been successful in other domains such as language processing, it suffers from inherently poor parallelization due to its sequential nature.
A final approach -- chosen in this work -- is to replace the matrix $\vec{F}$ with a trainable function $f^a_\theta(\vec{c}_{Ik})$, which computes the backflows based on some descriptor $\vec{c}_{Ik}$ of the orbital $k$ to be generated:
\begin{align*}
    &h_{i\nu} = h_\theta(\vec{r}_i, \lbrace\vec{r}\rbrace, \lbrace\vec{R}\rbrace, \lbrace\vec{Z}\rbrace)_\nu\\
    &\varphi^{d}_{\theta}(\vec{r}_i, \vec{R}_I, \vec{c}_{Ik}) = \exp\left(-|\vec{r}_i - \vec{R}_I|\,g_\theta^s(\vec{c}_{Ik})_d\right)\\
    &\Phi^d_{ik} =\sumions \varphi^{d}_{\theta}(\vec{r}_i, \vec{R}_I, \vec{c}_{Ik}) \sum_{\nu=1}^{\embdim} f_\theta^a(\vec{c}_{Ik})_{d\nu} h_{i\nu} 
\end{align*}
While there are several potential descriptors $\vec{c}_{Ik}$ for orbitals, one particularly natural choice is to use outputs of computationally cheap, conventional quantum chemistry methods such as Density Functional Theory or Hartree-Fock.
We compute orbital features based on the expansion coefficients of a Hartree-Fock calculation, by using orbital localization and a graph convolutional network (GCN), as outlined in Sec \ref{sec:localized_hf}. 
We then map these features to \emph{transferable atomic orbitals (TAOs)} $\Phi^d_{ik}$, using (anti-)symmetric functions $f^a_\theta$ and $g^s_\theta$ as illustrated in Fig. \ref{fig:architecture}.

\subsection{Properties of our ansatz}
\begin{figure*}
    \centering
    \includegraphics[width=\textwidth]{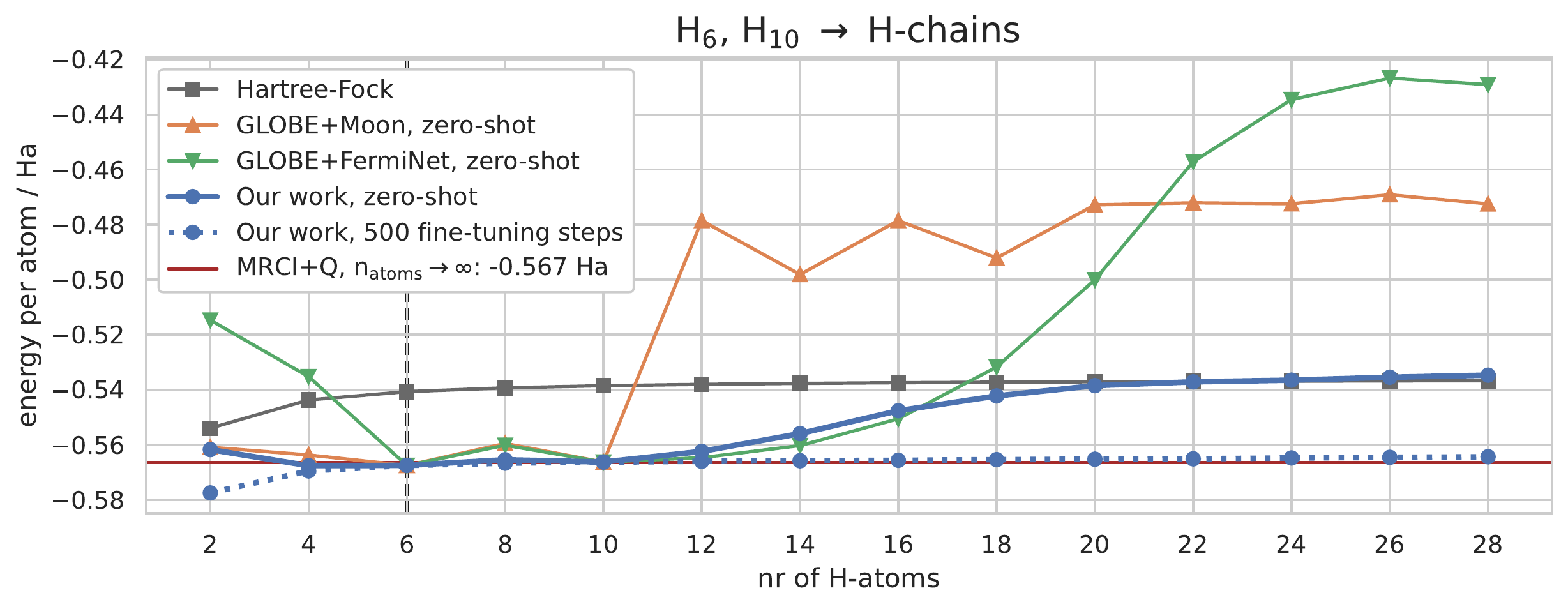}
    \caption{Zero-shot transferability of the ansatz to chemically similar, larger systems. While Moon cannot successfully transfer to larger chains, our ansatz successfully predicts zero-shot energies for up to 2x larger molecules and achieves very high accuracy with little pre-training.}
    \label{fig:size_consistency_HChains}
\end{figure*}
These TAOs fulfil many properties, which are desirable for a wavefunction ansatz: 
\begin{itemize}
    \item \textbf{Constant parameter count}: The number of parameters in the ansatz is independent of system size. In previous approaches \cite{hermannDeepneuralnetworkSolutionElectronic2020, pfauInitioSolutionManyelectron2020, scherbelaSolvingElectronicSchrodinger2022} the number of parameters grows with the number of particles, making it impossible to use a single ansatz across systems of different sizes. In particular backflows and envelope exponents have typically been chosen as trainable parameters of shape $[\norb \times \ndet]$. In our ansatz the backflows $\vec{F}$ are instead computed by a single function $f_\theta$ from multiple inputs $\vec{c}_{Ik}$.
    \item \textbf{Equivariant to sign of HF-orbital}: Orbitals of a HF-calculation are obtained as eigenvectors of a matrix and are thus determined only up to their sign (or their phase in the case of complex eigenvectors).
    We enforce that the functions $f^a_\theta$, $g^s_\theta$ are (anti-)symmetric with respect to $\vec{c}_{Ik}$. 
    Therefore our orbitals $\Phi^d_{ik}$ are equivariant to a flip in the sign of the HF-orbitals used as inputs: $\Phi(-\vec{c}_{Ik}) = -\Phi(\vec{c}_{Ik})$. 
    Therefore during supervised pre-training, the undetermined sign of the reference orbitals becomes irrelevant, leading to faster convergence as demonstrated in Sec. \ref{sec:results_hf_phase}. 
    \item \textbf{Locality}: When using localized HF-orbitals as input, the resulting TAOs are also localized. Localized HF-orbitals are orbitals which have non-zero orbital features $\vec{ \widetilde{c}}_{Ik}$ only on some subset of atoms. Since we enforce the backflow $f^a_{\theta}$ to be antisymmetric (and thus $f^a(\vec{0}) = \vec{0}$), the resulting TAOs have zero contribution from atoms $I$ with $\vec{c}_{Ik} = \vec{0}$.
    \item \textbf{High expressivity}: We empirically find that our ansatz is sufficiently expressive to model ground-state wavefunctions to high accuracy. This is in contrast to previous approaches which were based on incorporating ab-initio orbitals \cite{hermannDeepneuralnetworkSolutionElectronic2020}, which could not reach chemical accuracy even for small molecules.
\end{itemize}

\subsection{Size consistency of the ansatz}
\label{sec:size_consistency}
One design goal of the ansatz is to allow transfer of weights from small systems to larger systems. In particular, if a large system consists of many small previously seen fragments, one would hope to obtain an energy which corresponds approximately to the sum of the fragment energies.
One simple test case, are chains of equally spaced Hydrogen atoms of increasing lengths. These systems have been studied extensively using high-accuracy methods \cite{simonscollaborationonthemany-electronproblemSolutionManyElectronProblem2017}, because they are small systems which already show strong correlation and are thus challenging to solve.
We test our method by pre-training our ansatz on chains of length 6 and 10, and then evaluating the model (with and without subsequent fine-tuning) for chain lengths between 2 and 28.
Fig. \ref{fig:size_consistency_HChains} shows that our ansatz achieves very high zero-shot-accuracy in the interpolation regime ($n_\text{atoms}$ = 10) and for extrapolation to slightly larger or slightly smaller chains ($n_\text{atoms}$ = 4, 12). Even when extrapolating to systems of twice the size ($n_\text{atoms}$ = 20), our method still outperforms a Hartree-Fock calculation and eventually converges to an energy close the the Hartree-Fock solution. Fine-tuning the pre-trained model for only 500 steps, yields near perfect agreement with the specialized MRCI+Q method.

This good performance stands in stark contrast to other approaches such as GLOBE+\allowbreak FermiNet or GLOBE+\allowbreak Moon, studied in \cite{gaoGeneralizingNeuralWave2023}:
Both GLOBE-variants yield 5-6x higher errors in the interpolation regime and both converge to much higher energies for larger chains. 
While our approach yields Hartree-Fock-like energies for very long chains, GLOBE+\allowbreak FermiNet and GLOBE+\allowbreak Moon yield results that are outperformed even by assuming a chain of non-interacting H-atoms, which would yield an energy per atom of -0.5 Ha.
For modest extrapolations ($n=12$ to $n=20$) our zero-shot results yield 3 - 20x lower errors than GLOBE+Moon.

\subsection{Equivariance with respect to HF-phase}
\label{sec:results_hf_phase}
\begin{figure}[h!bt]
    \centering
    \includegraphics[width=\columnwidth]{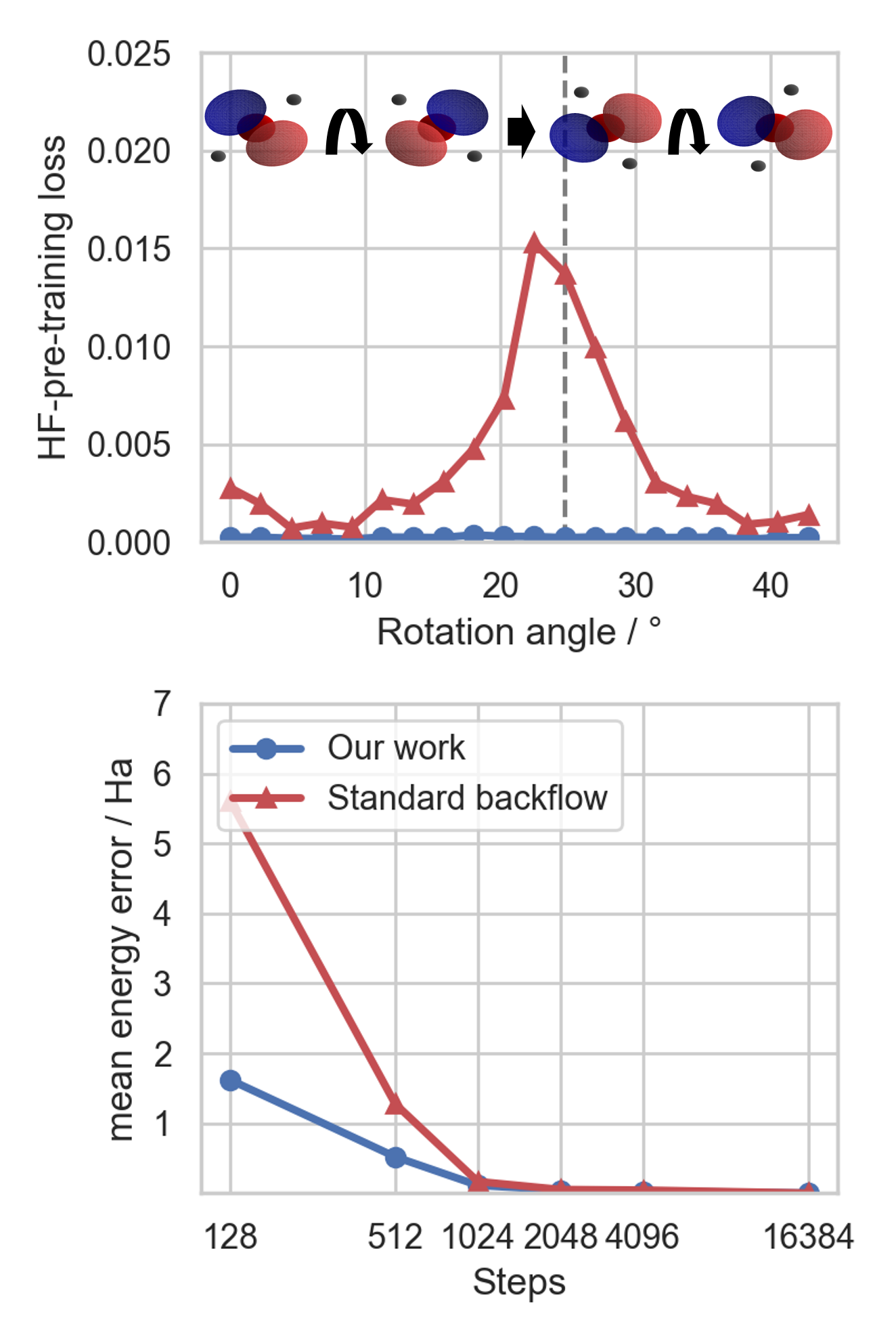}
    \caption{Accuracy when HF-pre-training against rotated H$_2$O molecules, which contain a change of sign in the Hartree-Fock-p-orbitals of the Oxygen atom. Comparing a shared optimization of a backflow-based neural network wavefunction (Standard backflow) against TAOs. Top: HF-pre-training loss averaged over 20 geometries and the last 100 samples for each rotation angle. Bottom: Mean energy error vs. reference, averaged across all geometries.}
    \label{fig:h20_phaseflip}
\end{figure}
Due to the (anti-)symmetrization of the TAOs, our orbitals are equivariant with respect to a change of sign of the Hartree Fock orbitals. 
Therefore, a sign change of the HF-orbitals during HF-pre-training has no effect on the optimization of the wavefunction. 
One test case to assure this behaviour is the rotation of a H$_2$O molecule, where we consider a set of 20 rotations of the same geometry, leading to a change of sign in the p-orbitals of the Oxygen atom (cf. Fig. \ref{fig:h20_phaseflip}). 
We evaluate our proposed architecture and compare it against a na\"ive approach, where we use a standard backflow matrix $\vec{F}$, instead of a trainable, anti-symmetrized function $f^a_\theta$. 
In Fig. \ref{fig:h20_phaseflip} we can see a clear spike in the HF-pre-training loss at the position of the sign flip for the standard backflow-type architecture, causing slower convergence during the subsequent variational optimization. 
Although, in this specific instance the orbital sign problem could also be overcome, without our approach by correcting the phase of each orbital to align them across geometries, phase alignment is not possible in all circumstances. For example, there are geometry trajectories, where the Berry phase prevents such solutions \cite{westermayr_machine_2021}. 

\subsection{Transfer to larger, chemically similar compounds}
\label{sec:reuse_from_smaller}
\begin{figure*}[h]
    \centering
    \includegraphics[width=\textwidth]{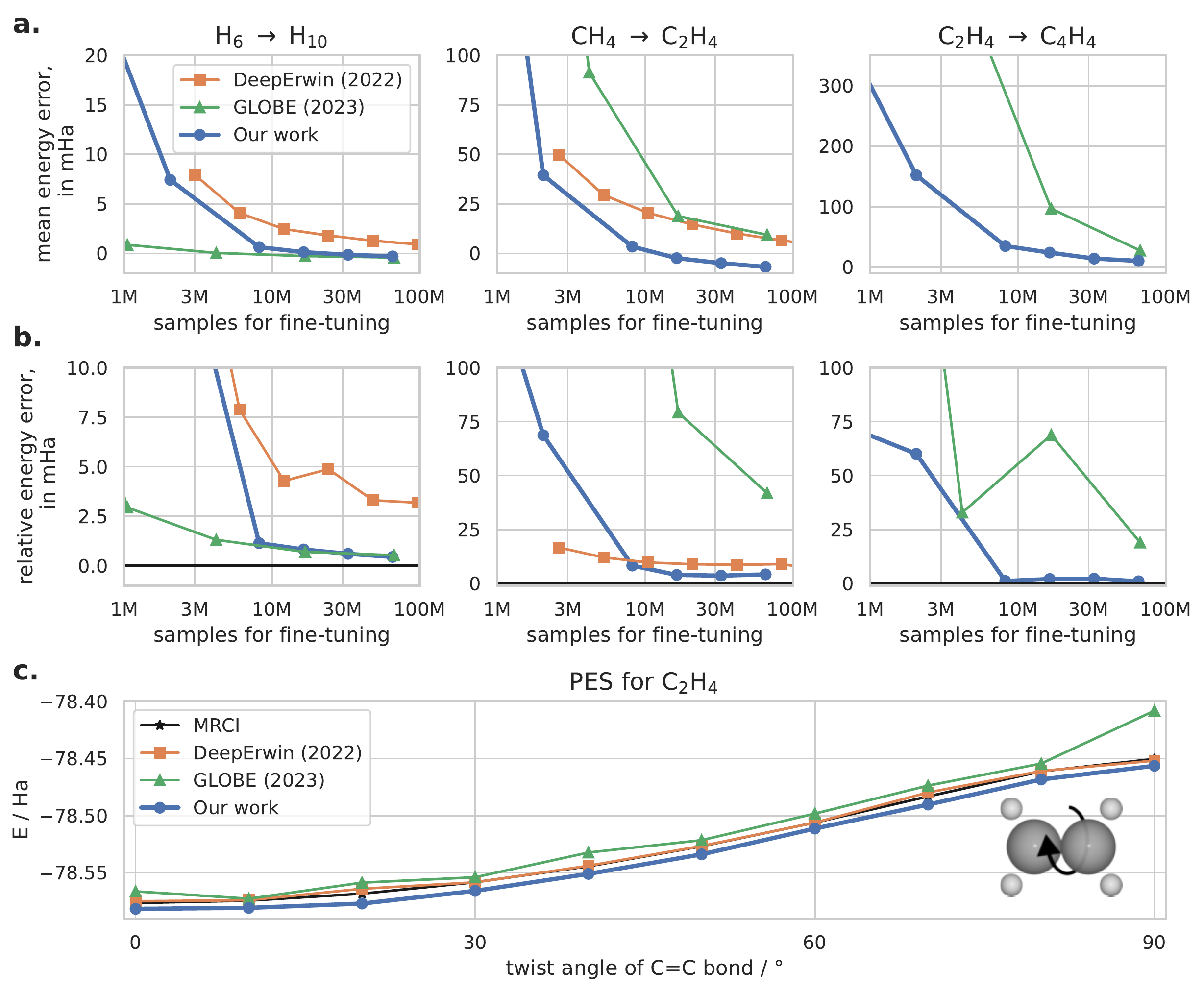}
    \caption{Accuracy when pre-training the model on small compounds and reusing it for larger compounds. Top: Mean energy error vs reference, averaged across all geometries of the test set. Middle: Error of relative energies, measured as $\max_g (E_g - E^\text{ref}_g) - \min_g (E_g - E^\text{ref}_g)$. Bottom: Final PES for the Ethene molecule for each method.}
    \label{fig:reuse_from_smaller}
\end{figure*}
To test the generalization and transferability of our approach, we perform the following experiment: First, we train our ansatz on a dataset of multiple geometries of a single, small compound (e.g. 20 distorted geometries of Methane). 
For this training, we follow the usual procedure of supervised HF-pre-training and subsequent variational optimization as outlined in Sec. \ref{sec:vmc}.
After 64k variational optimization steps, we then re-use the weights for different geometries of a larger compound (e.g. distorted geometries of Ethene). 
We fine-tune the model on this new geometry dataset for a variable number of steps and plot the resulting energy errors in Fig. \ref{fig:reuse_from_smaller}. We do not require supervised HF-pre-training on the new, larger dataset.
We perform this experiment for 3 pairs of test systems: Transferring from geometries of Hydrogen-chains with 6 atoms each, to chains with 10 atoms each, transferring from Methane to Ethene, and transferring from Ethene to Cyclobutadiene.

We compare our results to the earlier DeepErwin approach \cite{scherbelaSolvingElectronicSchrodinger2022}, which only partially reused weights,  and GLOBE, a concurrent pre-print \cite{gaoGeneralizingNeuralWave2023} which reuses all weights.
To measure accuracy we compare two important metrics: First, the mean energy error (averaged across all geometries $g$ of the test dataset) $\frac{1}{N}\sum_g (E_g - E^\text{ref}_g)$, which reflects the method's accuracy for absolute energies.
Second, the maximum relative energy error $\max_g (E_g - E^\text{ref}_g) - \min_g (E_g - E^\text{ref}_g)$, which reflects the method's consistency across a potential energy surface. 
Since different studies use different batch-sizes and different definitions of an epoch, we plot all results against the number of MCMC-samples used for variational optimization, which is very closely linked to computational cost. 

Compared to other approaches, we find that our method yields substantially lower and more consistent energies.
On the toy problem of H$_6$ to H$_{10}$ our approach and GLOBE reach the same accuracy, while DeepErwin converges to higher energies. 
For the actual molecules Ethene (C$_2$H$_4$) and Cyclobutadiene (C$_4$H$_4$) our approach reaches substantially lower energies and much more consistent potential energy surfaces. 
When inspecting the resulting Potential Energy Surface for Ethene, we find that we obtain qualitatively similar results as DeepErwin, but obtain energies that are $\approx$\,6\,mHa lower (and thus more accurate).
GLOBE on the other hand does not yield a qualitatively correct PES for this electronically challenging problem. 
It overestimates the energy barrier at 90$^\circ$ twist angle by $\approx$\,50\,mHa and yields a spurious local minimum at 10$^\circ$.
We observe similar results on the Cyclobutadiene geometries, where our approach yields energy differences that are in close agreement to the reference energies, while the GLOBE-results overestimate the energy difference by $\approx$\,20\,mHa.

\subsection{Towards a first foundation model for neural network wavefunctions}
\label{sec:foundation_model}
While the experiments in Sec. \ref{sec:reuse_from_smaller} demonstrate the ability to pre-train our model and fine-tune it on a new system, the resulting pre-trained models are of little practical use, since they are only pre-trained on a single compound each and can thus not be expected to generalize to chemically different systems.
To obtain a more diverse pre-training dataset, we compiled a dataset of 360 distorted geometries, spread across 18 different compounds. 
The dataset effectively enumerates all chemically plausible molecules with up to 18 electrons containing the elements H, C, N, and O. For details on the data generation see \ref{appendix:tinymol_dataset}.
We pre-train a base-model for 500,000 steps on this diverse dataset and subsequently evaluate its performance, when computing Potential Energy Surfaces. 
We evaluate its error both for compounds that were in the pre-training dataset (with different geometries), as well as for new, larger, out-of-distribution compounds which were not present in the pre-training dataset.
We compare the results against a baseline model, which uses the default method of supervised HF-pre-training and subsequent variational optimization.

\begin{figure}
    \centering
    \includegraphics[width=0.5\textwidth]{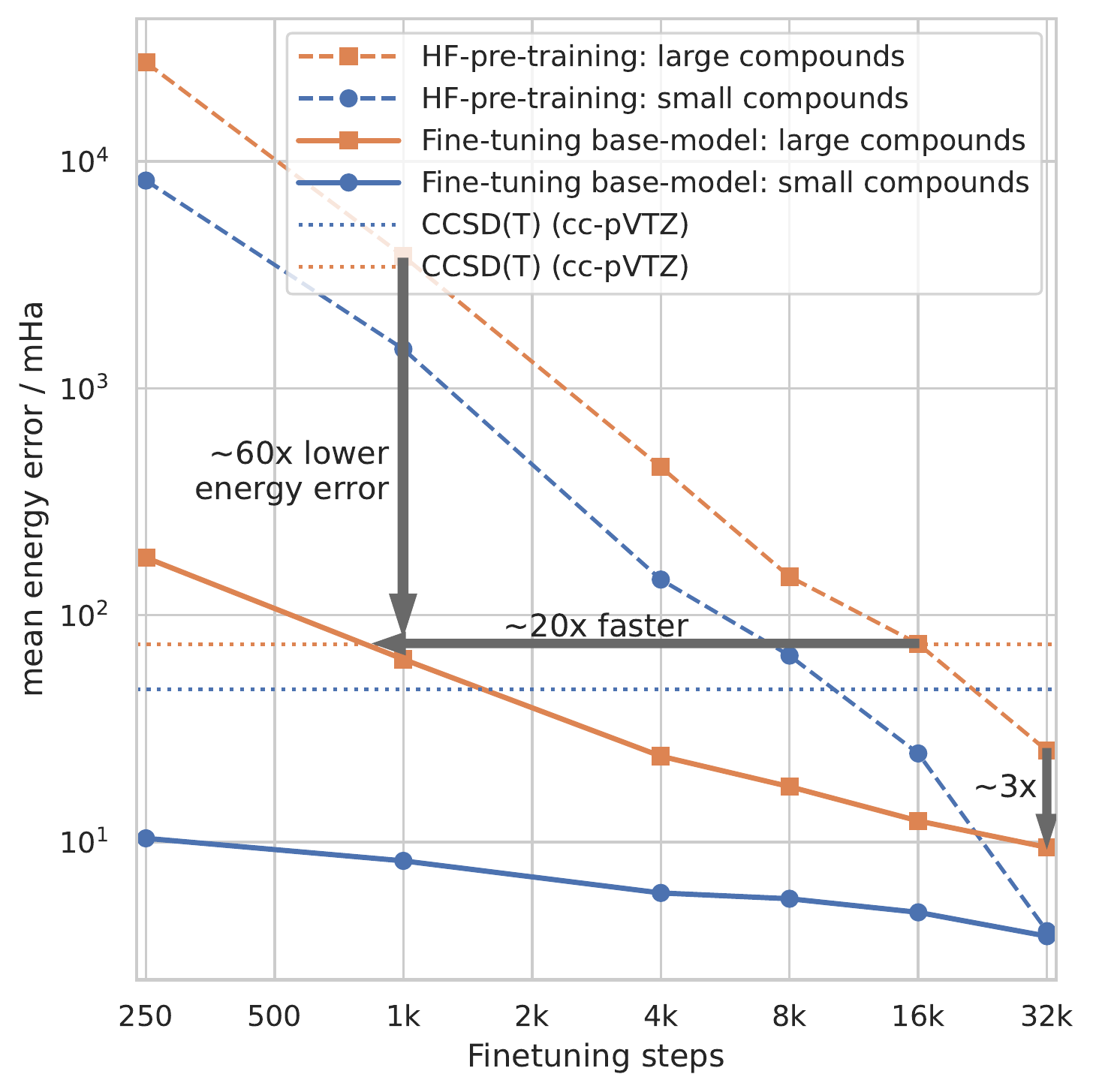}
    \caption{Fine-tuning of pre-trained foundation model (solid lines) vs. fine-tuning Hartree-Fock-pre-trained models (dashed lines) for 70 different geometries. Blue: Small compounds, with geometries similar to geometries in pre-training dataset. Orange: Larger compounds outside pre-training dataset.}
    \label{fig:evaluation_of_foundational_model}
\end{figure}

Fig. \ref{fig:evaluation_of_foundational_model} shows that fine-tuning this pre-trained model yields substantially lower energies than the usual optimization from a HF-pre-trained model. 
For example, when optimizing for 8k steps, we obtain 8x lower energy errors for large out-of-distribution compounds, and 12x lower energies for small in-distribution compounds.
When evaluating the model for up to 32k steps, we find that for small molecules both approaches converge to the same energy.
For large molecules the final energy error obtained by fine-tuning the base-model is 3x lower than optimization of a HF-pre-trained model.

\subsection{Scaling behaviour}
\label{sec:scaling_behaviour}
\begin{figure}
    \centering
    \includegraphics[width=\columnwidth]{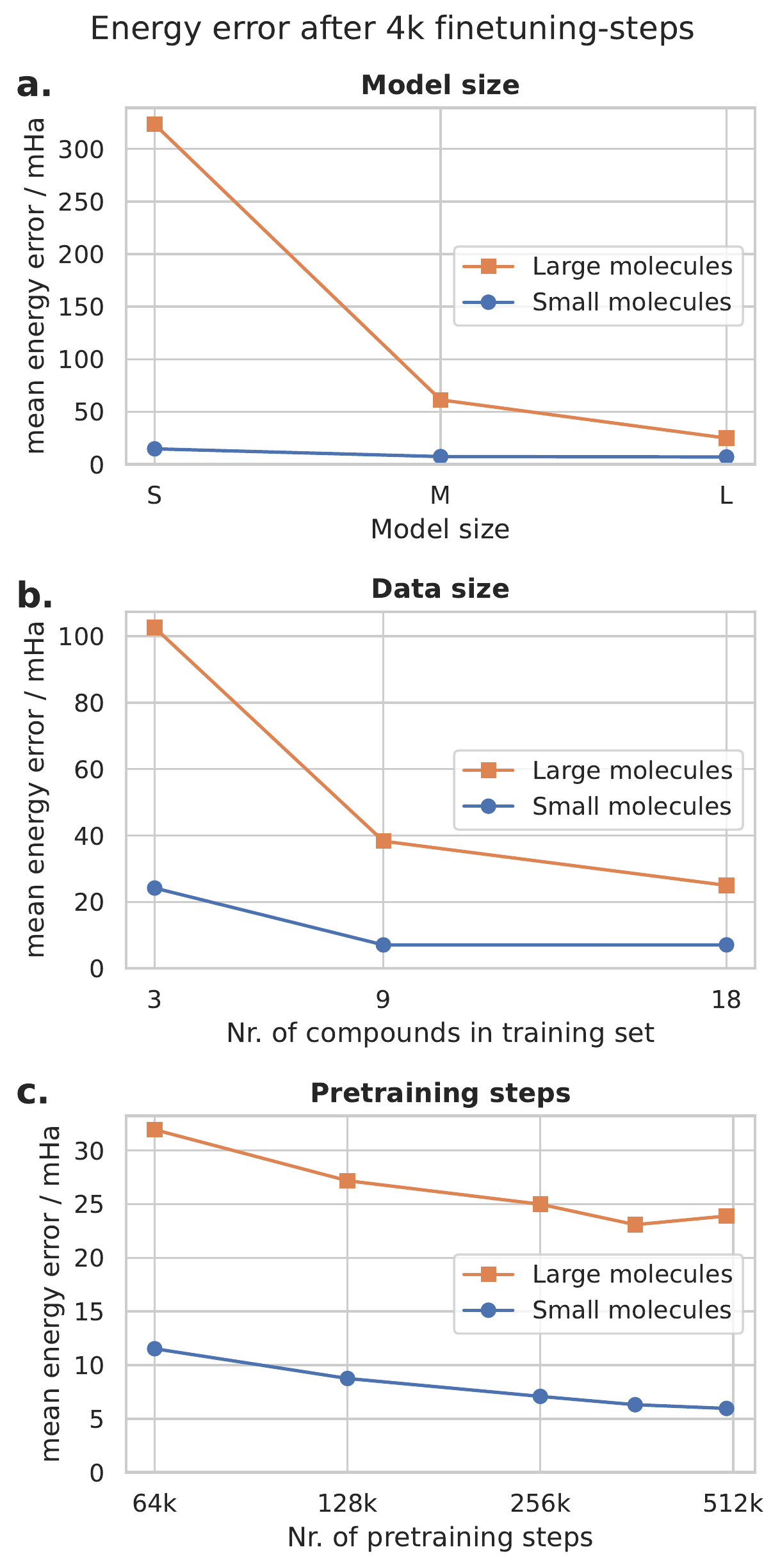}
    \caption{Error when fine-tuning the pre-trained model for 4000 steps on small in-distribution geometries and larger out-of-distribution geometries.}
    \label{fig:ablation_of_foundational_model}
\end{figure}

In many domains, increasing the amount of pre-training, has led to substantially better results, even without qualitative changes to the architecture \cite{hoffmannTrainingComputeOptimalLarge2022}.
To investigate the scalability of our approach, we vary the three key choices, along which one could increase the scale of pre-training:
The size of the wavefunction model, the number of compounds and geometries present in the pre-training-dataset, and the number of pre-training steps.
Starting from a large model, trained on 18x20 geometries, for 256k pre-training steps, we independently vary each parameter. 
We test 3 different architectures sizes, with decreasing layer width and depth for the networks $f_\theta$, $g_\theta$, and $\text{GCN}_\theta$ (cf. \ref{appendix:computational_settings}).
We test 3 different training sets, with decreasing number of compounds in the training set, with 20 geometries each (cf. \ref{appendix:tinymol_dataset}).
Finally, we evaluate model-checkpoints at different amounts of pre-training, ranging from 64k steps to 512k steps.
Fig. \ref{fig:ablation_of_foundational_model} depicts the accuracy obtained by subsequently fine-tuning the resulting model for just 4000 steps on the evaluation set.  
In each case, increasing the scale of pre-training clearly improves evaluation results -- both for the small in-distribution compounds, as well as the larger out-of-distribution compounds.
We find a strong dependence of the accuracy on the model size and number of compounds in the pre-training dataset, and a weaker dependency on the number of pre-training steps.
While our computational resources, currently prohibit us from training at larger scale, the results indicate that our approach may already be sufficient to train an accurate multi-compound, multi-geometry foundation model for  wavefunctions.

\section{Discussion}
\label{sec:discussion}

This work presents an ansatz for deep-learning-based VMC, which can in principle be applied to molecules of arbitrary size.
We demonstrate the favourable properties of our ansatz, such as extensivity, zero-shot prediction of wavefunctions for similar molecules (Sec. \ref{sec:size_consistency}), invariance to the phase of orbitals (Sec. \ref{sec:results_hf_phase}) and fast finetuning for larger, new molecules (Sec. \ref{sec:reuse_from_smaller}).
Most importantly, Sec. \ref{sec:foundation_model} is, to our knowledge, the first successful demonstration of a general wavefunction, which has successfully been trained on a diverse dataset of compounds and geometries.
We demonstrate that the dominating deep-learning paradigm of the last years -- pre-training on large data and fine-tuning on specific problems -- can also be applied to the difficult problem of wavefunctions.
While previous attempts \cite{scherbelaSolvingElectronicSchrodinger2022, gaoGeneralizingNeuralWave2023} have failed to obtain high-accuracy energies from pre-trained neural network wavefunctions, we find that our approach yields accurate energies and does so at a fraction of the cost needed without pre-training.
We furthermore demonstrate in Sec. \ref{sec:scaling_behaviour} that results can be improved systematically by scaling up any aspect of the pre-training: Model size, data-size, or pre-training-steps. 

Despite these promising results, there are many open questions and limitations which should be addressed in future work. 
First, we find that our ansatz currently does not fully match the accuracy of state-of-the-art single-geometry DL-VMC ansätze. 
While our approach consistently outperforms conventional variational methods such as MRCI or CCSD at finite basis set, larger, computationally more expensive DL-VMC models can reach even lower energies.
Exchanging our message-passing-based electron-embedding, with recent attention based approaches \cite{vonglehnSelfAttentionAnsatzAbinitio2022} should lead to higher accuracy.
Furthermore we have made several deliberate design choices, which each trade-off expressivity (and thus potentially accuracy) for computational cost: 
We do not exchange information across orbitals and we base our orbitals on computationally cheap HF-calculations. 
Including attention or message passing across orbitals (e.g. similar to \cite{gaoGeneralizingNeuralWave2023}), and substituting HF for a trainable, deep-learning-based model should further increase expressivity.
While we currently use HF-orbitals due to their widespread use and low computational cost, our method does not rely on a specific orbital descriptor.
We could substitute HF for a separate model such as PhysNet \cite{unkeSEEquivariantPrediction2021} or SchnOrb \cite{gasteggerDeepNeuralNetwork2020} to compute orbital descriptors $\vec{c}_{Ik}$, leading to a fully end-to-end machine-learned wavefunction.
Second, while we include useful physical priors such as locality, we do not yet currently use the invariance of the Hamiltonian with respect to rotations, inversions or spin-flip.
E3-equivariant networks have been highly successful for neural network force-fields, but have not yet been applied to wavefunctions due to the hitherto unsolved problem of symmetry breaking \cite{gaoAbInitioPotentialEnergy2021}. 
Using HF-orbitals as symmetry breakers, could open a direct avenue towards E3-equivariant neural network wavefunctions.
Third, while we use locality of our orbitals as a useful prior, we do not yet use it to reduce computational cost. 
By enforcing sparsity of the localized HF-coefficients, one could limit the evaluation of orbitals to a few participating atoms, instead of all atoms in the molecule. 
While the concurrent GLOBE approach enforces its orbitals to be localized at a single position, our approach naturally lends itself to force localization at a given number of atoms, allowing for a deliberate trade-off of accuracy vs. computational cost.
Lastly, we observe that our method performs substantially better, when dedicating more computational resources to the pre-training. 
While we are currently constrained in the amount of computational resources, we hope that future work will be able to scale up our approach.
To facilitate this effort we open source our code, dataset as well as model parameters.

\section{Methods}
\subsection{Variational Monte Carlo}
\label{sec:vmc}
Considering the Born-Oppenheimer approximation, a molecule with $\nel$ electrons and $\nnuc$ nuclei can be described by the time-independent Schrödinger equation 
\begin{equation}
    \hat{H} \psi = E\psi
\label{eq:schroedinger_equation}
\end{equation}
with the Hamiltonian 
\begin{align}
    \hat{H} =& -\frac{1}{2}\sum_{i} \nabla^2_{\vec{r}_i} + \sum_{i>j} \frac{1}{|{\vec{r}_i - \vec{r}_j}|} \nonumber \\ 
    &+ \sum_{I>J}\frac{Z_I Z_J}{|{\vec{R}_I - \vec{R}_J}|} 
    - \sum_{i,I} \frac{Z_I}{|{\vec{r}_i - \vec{R}_I}|}
\end{align}
By $\vec{r} = (\vec{r}_1, \dots, \vec{r}_{n_{\uparrow}}, \dots, \vec{r}_{\nel}) \in \mathbb{R}^{3 \times \nel}$ we denote the set of electron positions divided into $n_{\uparrow}$ spin-up and $n_{\downarrow}$ spin-down electrons. For the coordinates and charges of the nuclei we write $\vec{R}_I$, $Z_I$, with $I \in \lbrace 1, \dots, \nnuc \rbrace$. The solution to the electronic Schrödinger equation $\psi$ needs to fulfill  the anti-symmetry property, i.e. $\psi( \mathcal{P} \vec{r}) = - \psi(\vec{r})$ for any permutation $ \mathcal{P} $  of two electrons of the same spin. 
Finding the groundstate wavefunction of a system, corresponds to finding the solution to Eq. \ref{eq:schroedinger_equation}, with the lowest eigenvalue $E_0$. Using the Rayleigh-Ritz principle, an approximate solution can be found through minimization of the loss
\begin{align}
    \mathcal{L}(\psi_{\theta}) = \mathbb{E}_{\vec{r}\sim \psi_\theta^2(\vec{r})} 
    \left[\frac{(\hat{H}\psi_\theta)(\vec{r})}{\psi_\theta(\vec{r})} \right] \geq E_0,
\label{eq:loss}
\end{align}
using a parameterized trial wavefunction $\psi_\theta$.
The expectation value in Eq. \ref{eq:loss} is computed by drawing samples $\vec{r}$ from the unnormalized probability distribution $\psi_\theta^2(\vec{r})$ using Markov Chain Monte Carlo (MCMC). The effect of the Hamiltonian on the wavefunction can be computed using automatic differentiation and the loss is minimized using gradient based minimization.
A full calculation typically consists of three steps:
\begin{enumerate}
    \item \textbf{Supervised HF-pre-training}: Minimization of the difference between the neural network ansatz and a reference wavefunction (e.g. a Hartree-Fock calculation) $||\psi_\theta - \psi^\hf||$. This is the only part of the procedure which requires reference data, and ensures that the initial wavefunction roughly resembles the true groundstate. While this step is in principle not required, it substantially improves thes stability of the subsequent variational optimization.
    \item \textbf{Variational optimization}: Minimization of the energy (Eq. \ref{eq:loss}) by drawing samples from the wavefunction using MCMC, and optimizing the parameters $\theta$ of the ansatz using gradient based optimization.
    \item \textbf{Evaluation}: Evaluation of the energy by evaluating Eq. \ref{eq:loss} without updating the parameters $\theta$, to obtain unbiased estimates of the energy.
\end{enumerate}
To obtain a single wavefunction for a dataset of multiple geometries or compounds, only minimal changes are required. 
During supervised and variational optimization, for each gradient step we pick one geometry from the dataset.
We pick geometries either in a round-robin fashion, or based on the last computed energy variance for that geometry.
We run the Metropolis Hastings algorithm \cite{hastingsMonteCarloSampling1970} for that geometry to draw electron positions $\vec{r}$ and then evaluate energies and gradients.
For each geometry we keep a distinct set of electron samples $\vec{r}$.

\subsection{Obtaining orbital descriptors from Hartree-Fock}
\label{sec:localized_hf}
As discussed in Sec. \ref{sec:extending_ansatz_to_multi_compound}, our ansatz effectively maps uncorrelated, low-accuracy Hartree-Fock orbitals, to correlated, high-accuracy neural network orbitals.
The first step in this approach is to obtain orbital descriptors $\vec{c}_k$ for each orbital $k$, based on a Hartree-Fock calculation.

The Hartree-Fock method uses a single determinant as ansatz, composed of single-particle orbitals $\phi_k$:
\begin{align}
    \psi^\hf(\vec{r}_1, \dots, \vec{r}_\nel) =& \det \left[\Phi^\hf_{ik}\right]_{i,k=1\dots \nel} \\  \Phi^\hf_{ik} :=& \phi^\hf_k(\vec{r}_i)
\end{align}
For molecules, these orbitals are typically expanded in atom-centered basis-functions $\mu(\vec{r})$, with $\nbasis$ functions centered on each atom $I$:
\begin{equation}
    \phi^\text{HF}_k(\vec{r}) = \sumions \sum_{b=1}^\nbasis \alpha_{k,Ib} \; \mu_b(\vec{r}-\vec{R}_I),
\end{equation}
The coefficients $\vec{\alpha}_k$ and the corresponding orbitals $\phi^\hf_k(\vec{r})$ are obtained as solutions of an eigenvalue problem and are typically delocalized, i.e. they have non-zero contributions from many atoms.
However, since $\det[U\Phi] = \det[U] \det[\Phi]$, the wavefunction is invariant under linear combination of orbitals by a matrix $U$ with $\det[U] = 1$. One can thus choose orbital expansion coefficients
\begin{equation}
    \widetilde{\alpha}_{k, Ib} = \sum_{k'=1}^\norb  \alpha_{k,Ib} U_{kk'}
\end{equation}
corresponding to orbitals which are maximally localized according to some metric.
Several different metrics and corresponding localization schemes,  such as Foster-Boys \cite{fosterCanonicalConfigurationalInteraction1960} or Pipek-Mezey \cite{pipekFastIntrinsicLocalization1989}, have been proposed and are easily available as post-processing options in quantum chemistry codes.
We use the Foster-Boys method as implemented in pySCF \cite{sunRecentDevelopmentsPySCF2020}.

Due to the fundamentally local nature of atom-wise orbital coefficients $\vec{\widetilde{\alpha}}_{Ik}$, which can be insufficient to distinguish orbitals, we use a fully connected graph convolutional neural network (GCN) to add context about the surrounding atoms.
We interpret each atom as a node (with node features $\vec{\widetilde{\alpha}}_{Ik}$) and the 3D inter-atomic distance vector $\vec{R}_{IJ}$ as edge features:
\begin{equation*}
    \vec{c}_k = \text{GCN}_\theta\left(\{\vec{\widetilde{\alpha}}_{Ik}\}_{I=1\dots\nnuc}, \{\vec{R}_{IJ}\}_{I,J=1\dots\nnuc}\right)
\end{equation*}
We embed the edge features using a cartesian product of Gaussian basis functions of the distance $R_{IJ}$ and the concatenation of the 3D-distance vector with the constant 1:
\begin{align*}
    &\vec{\widetilde{e}}_{IJ} = \exp\left(-\frac{(R_{IJ}-\mu_n)^2}{2\sigma_n^2}\right) \otimes \left[1 | \vec{R}_{IJ}\right]\\
    &\vec{e}^0_{IJ} = \text{MLP}(\vec{\widetilde{e}}_{IJ})\\
    &\vec{c}^0_{I} = \vec{\alpha}_I
\end{align*}
Each layer $l$ of the GCN consist of the following update rules
\begin{align*}
    &\vec{u}^{l}_{Ik} = \sum_J \vec{c}_{Jk}^l \odot \left(\vec{W}^l_g \vec{e}^l_{IJ}\right),\\
    &\vec{c}^{l+1}_{Ik} = \sigma(\vec{W}^l_v \vec{c}_{Ik} + \vec{W}^l_u \vec{u}^l_{Ik}), 
\end{align*}
with trainable weight matrices $\vec{W}^l_g$, $\vec{W}^l_v$, $\vec{W}^l_u$ and the SiLU activation function $\sigma$.
After $L$ iterations we use the final outputs as orbitals features:
\begin{equation*}
    \vec{c}_{Ik} := \vec{c}^L_{Ik}
\end{equation*}

\subsection{Mapping orbital descriptors to wavefunctions}
\label{sec:our_wf_ansatz}
To obtain entry $\Phi_{ik}$ of the Slater determinant, we combine a high-dimensional electron embedding $\vec{h}_i$ with a function of the orbital descriptor $\vec{c}_{Ik}$:
\begin{align*}
    &h_{i\nu} = h_\theta(\vec{r}_i, \lbrace\vec{r}\rbrace, \lbrace\vec{R}\rbrace, \lbrace\vec{Z}\rbrace)_\nu\\
    &\varphi^{d}_{\theta}(\vec{r}_i, \vec{R}_I, \vec{c}_{Ik}) = \exp\left(-|\vec{r}_i - \vec{R}_I|\,g_\theta^s(\vec{c}_{Ik})_d\right)\\
    &\Phi^d_{ik} =\sumions \varphi^{d}_{\theta}(\vec{r}_i, \vec{R}_I, \vec{c}_{Ik}) \sum_{\nu=1}^{\embdim} f_\theta^a(\vec{c}_{Ik})_{d\nu} h_{i\nu} 
\end{align*}
The functions $\text{GCN}^a_\theta$, $f_\theta^a$, and $g_\theta^s$ are trainable functions, which are enforced to be (anti-)symmetric with respect to change in sign of their argument $\vec{c}$:
\begin{align*}
    &\text{Symmetric $g^s_\theta$:}\\ 
    &g^s_\theta(\vec{c}) := g_\theta(\vec{c}) + g_\theta(-\vec{c})\\
    &\text{Antisymm. $f^a_\theta$:}\\
    &f^a_\theta(\vec{c}) := f_\theta(\vec{c}) - f_\theta(-\vec{c})\\
    &\text{Antisymm. GCN$^a_\theta$:}\\
    &\text{GCN}^a_\theta(\vec{\alpha}, \vec{R}) := \text{GCN}_\theta(\vec{\alpha}, \vec{R}) - \text{GCN}_\theta(-\vec{\alpha}, \vec{R})
\end{align*}
To obtain electron embeddings $\vec{h}_i$ we use the message-passing architecture outlined in \cite{gerardGoldstandardSolutionsSchrodinger2022}, which is invariant with respect to permutation of electrons of the same spin, or the permutation of ions.
\begin{align*}
    &\mathcal{G} = \{(\vec{R}_I, Z_I)\}_{I=1\dots\nnuc}\\
    &\mathcal{E}_\uparrow = \{\vec{r}_i\}_{i=1\dots\nup}\\
    &\mathcal{E}_\downarrow = \{\vec{r}_i\}_{i=\nup+1\dots\nel}\\
    &\vec{h}_i = h_\theta^\text{embed}(\vec{r}_i, \mathcal{G}, \mathcal{E}_\uparrow, \mathcal{E}_\downarrow)
\end{align*}
Note that during training, all samples in a batch come from the same geometry, and thus have the same values for $\vec{R}$, $\vec{Z}$, and $\vec{\widetilde{ \alpha}}$. 
While the embedding network $h^\text{embed}_\theta$, needs to be re-evaluated for every sample, the networks $\text{GCN}_\theta$, $f_\theta$, and $g_\theta$ only need to be evaluated once per batch, substantially reducing their impact on computational cost.
\section*{Acknowledgements}
We gratefully acknowledge financial support from the following grants: Austrian Science Fund FWF Project I 3403 (P.G.), WWTF-ICT19-041 (L.G.). The computational results have been achieved using the Vienna Scientific Cluster (VSC). The funders had no role in study design, data collection and analysis, decision to publish or preparation of the manuscript. Additionally, we thank Nicholas Gao for providing his results and data, Ruard van Workum for initial work on the python implementation for multi-compound optimization and Jan Hermann for fruitful discussions.

\section*{Author contributions}
MS, LG, and PG conceived the overall idea. MS conceived and implemented the ansatz, built the dataset and designed the experiments. 
LG gave input on the ansatz and worked on implementation. MS and LG performed the experiments.
MS and LG wrote the manuscript with input, supervision and funding from PG.
\newpage
\printbibliography

\newpage
\appendix
\section{Dataset used for pre-training of foundation model}
\label{appendix:tinymol_dataset}
\begin{figure*}
    \centering
    \includegraphics[width=0.95\textwidth]{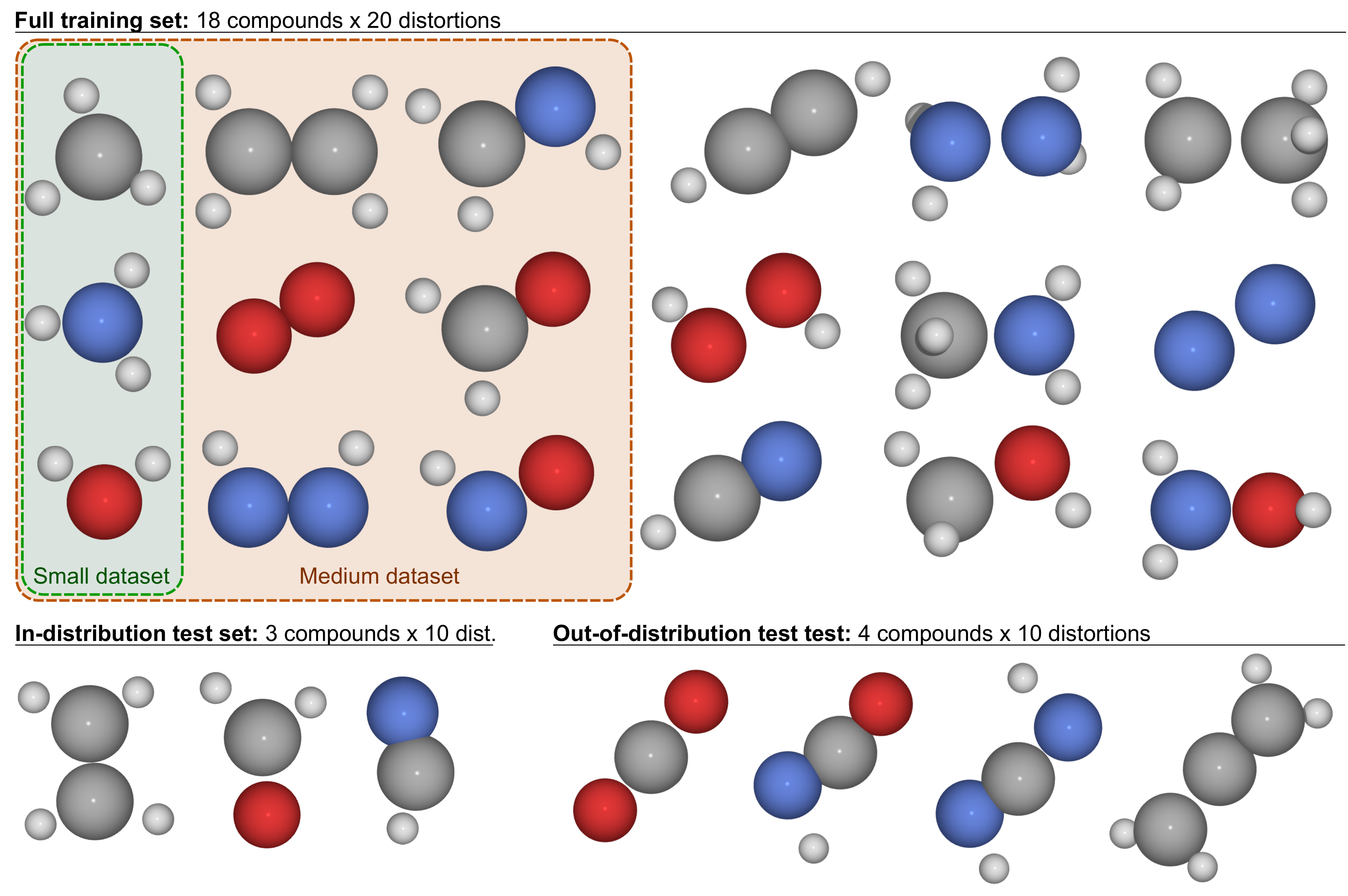}
    \caption{Compounds used for pre-training and evaluation of our model. Atom colors follow the usual convention of H = white, C = gray, N = blue, O = red. Note that the full training and test sets, each contain several distortions of each molecular geometry.}
    \label{fig:dataset_render}
\end{figure*}

We use RDKit \cite{RDKit} to generate all valid SMILES of molecules containing 1-3 atoms of the elements C, N, O. For each bond between atoms we allow single, double, and triple bonds. After saturating the molecules with Hydrogen, we perform force-field based Geometry relaxation using RDKit.
We obtain 18 compounds with 10-18 electrons, which we use for pre-training (cf. Fig. \ref{fig:dataset_render}) and 35 compounds with 20-24 electrons, of which we use some for evaluation. Contrary to other datasets of small molecules such as GDB-7, our dataset also includes compounds which do not contain Carbon, such as the nitrogen dimer N$_2$ or hydrogen peroxide H$_2$O$_2$.
To obtain a more diverse dataset we perturb each equilibrium geometry by applying Gaussian noise to the 3D coordinates. Since this can generate nonphysical geometries, we keep only geometries in which the perturbed inter-atomic distances are between 90\% - 140\% of the unperturbed distances.

\section{Reference energies}
Reference energies for H$_2$O in Fig. \ref{fig:h20_phaseflip} were computed using DL-VMC for 100,000 steps \cite{gerardGoldstandardSolutionsSchrodinger2022}.
Reference energies for H$_{10}$ and C$_2$H$_4$ in Fig. \ref{fig:reuse_from_smaller} were computed using MRCI-F12(Q) \cite{scherbelaSolvingElectronicSchrodinger2022}. 
Reference energies for C$_4$H$_4$ in Fig. \ref{fig:reuse_from_smaller} were computed using DL-VMC \cite{gao-pesnet++_2022}.
To compute reference energies for our dataset in Sec. \ref{sec:foundation_model}-\ref{sec:scaling_behaviour}, we used pySCF \cite{sunRecentDevelopmentsPySCF2020} to perform CCSD(T) calculations using the cc-pCVXZ basis sets.
We computed Hartree-Fock energies $E^\hf_X$ using basis-sets of valence $X=\{2,3,4\}$ and CCSD(T) energies $E^\text{CCSD(T)}_X$ using valence $X=\{2,3\}$.
To extrapolate to the complete-basis-set-limit, we followed \cite{pfauInitioSolutionManyelectron2020} and fitted the following functions with free parameters $E^\hf_\text{CBS}, E^\text{corr}_\text{CBS}, a, b, c$:
\begin{align*}
    &E^\hf_X = E^\hf_\text{CBS} + a e^{-bX} \\
    &E^\text{corr}_X := E^\hf_X - E^\text{CCSD(T)}_X = E^\text{corr}_\text{CBS} + c X^{-3} \\
    &E^\text{CCSD(T)}_\text{CBS} = E^\hf_\text{CBS} + E^\text{corr}_\text{CBS}
\end{align*}
While the CCSD energies are variational (and thus an upper bound to the true groundstate energy), neither the perturbative (T) treatment nor the CBS extrapolation are variational and thus the obtained reference energies may underestimate the true groundstate energies.

\section{Computational settings}
\label{appendix:computational_settings}
For a more detailed summary and explanation of the high-dimensional embedding structure we refer to the original work \cite{gerardGoldstandardSolutionsSchrodinger2022}. In all experiments we relied on the second order optimizer K-FAC  \cite{martens2015optimizing, kfac_repo_jax}. For the foundation model in section \ref{sec:foundation_model} we increased the initial damping by $10$x and decreased it to $1 \times 10^{-3}$ with a inverse scheduler. All runs reusing pre-trained weights used a $5$x lower initial learning rate with a learning rate scheduler offset $o = 32,000$, i.e. $\text{lr}(t) = \text{lr}_0 (1+(t+o)/6000)^{-1}$. All pre-trained runs in section \ref{sec:reuse_from_smaller} used $64,000$ optimization steps. 
The foundation model in section \ref{sec:foundation_model} used $512,000$ optimization steps due to the larger and more diverse training corpus.

The small- and medium-sized model for our ablation study in Fig. \ref{fig:ablation_of_foundational_model} differ from the large model by the number of hidden layers for $f_\theta$ and $g_\theta$, the number of neurons per layer, and the number of iterations of the GCN:
The small model uses no hidden layers and no GCN. The medium-sized model uses one hidden layer of width 64 for $g_\theta$ and 128 for $f_\theta$, and one iteration of the GCN.
The small, medium and large models respectively have 0.8 mio, 1.2 mio. and 2.0 mio parameters.
\begin{table*}[!b]
\caption{Hyperparameter settings used in this work \label{table:hyperparams}}
\begin{tabularx}{\textwidth}{lXc}
\toprule
\multirow{2}{*}{
\textbf{HF-pre-training}
}
		 &Pre-training basis set & 6-31G + p-functions for H\\
		 &Pre-training steps per geometry & 100-500\\
\midrule
\multirow{4}{*}{ \textbf{Embedding}}
		 &Hidden dimension of $\vec{h}_i$      & 256\\ 
		 &Dimension of SchNet convolution       & 32\\ 
		 &\textnumero\, iterations embedding		                  & 4\\ 
		 &Activation function & tanh\\
\midrule
\multirow{11}{*}{\shortstack[l]{
\textbf{Transferable}\\
\textbf{atomic orbitals}
}}
		 & \textnumero\, determinants $\ndet$   & 4\\ 
          & Basis set & 6-31G + p-functions for H \\
          & \textnumero\, hidden layers $f_\theta$ & 2\\
          & Hidden dimension of $f_\theta$      & 256\\     
          & \textnumero\, hidden layers $g_\theta$ & 2\\
          & \textnumero\, hidden dimension $g_\theta$      & 128\\   
          & \textnumero\, iterations GCN  &  2 \\
          & \textnumero\, Gaussian basis functions & 16 \\
          & Hidden edge embedding dimension & 32 \\
          & Hidden node embedding dim. & 16 \\ 
		 & Activation function & SiLU\\
\midrule
\multirow{3}{*}{\shortstack[l]{
\textbf{Markov Chain}\\
\textbf{Monte Carlo}
}}
		 &\textnumero\, walkers                        & $2048$	\\
		 &\textnumero\, decorrelation steps            & 20	\\
		 &Target acceptance prob.         & 50\%	\\
\midrule
\multirow{7}{*}{\shortstack[l]{
\textbf{Variational}\\
\textbf{pre-training}
}}
		 &Optimizer		                        & KFAC\\
		 &Damping 	                            & $1 \times 10^{-3}$ 	\\
		 &Norm constraint	                    & $3 \times 10^{-3}$ 	\\
		 &Batch size 	                        & $2048$	\\
		 &Initial learning rate $\text{lr}_0$	& $0.1$	\\
		 &Learning rate decay   	            & $\text{lr}(t) = \text{lr}_0 (1+t/6000)^{-1}$ 	\\
		 &Optimization steps                    &64,000 - 512,000\\
\midrule
\multirow{2}{*}{\shortstack[l]{
\textbf{Changes for}\\
\textbf{fine-tuning}
}}
		 &Learning rate decay   	            & $\text{lr}(t) = \text{lr}_0 (7+t/6000)^{-1}$ 	\\
		 &Optimization steps                    &0 - 32,000\\
\bottomrule
\end{tabularx}
\end{table*}

\end{document}